\documentclass[journal=jacsat,manuscript=article]{achemso}

\usepackage{chemformula} 
\usepackage[T1]{fontenc} 
\usepackage{float}
\usepackage{siunitx}
\usepackage{enumitem}
\usepackage{amsfonts}
\usepackage{setspace}
\usepackage{hyperref}

\author{Sandro L. Camenzind}
\altaffiliation{These authors contributed equally to this work}
\email{casandro@phys.ethz.ch}

\author{Lukas Lang}
\altaffiliation{These authors contributed equally to this work}

\author{Benjamin Willenberg}
\author{Justinas Pupeikis}
\author{Hayk Soghomonyan}

\affiliation[ETH Physics]
{\onehalfspacing Department of Physics, Institute for Quantum Electronics, ETH Zurich, Switzerland}

\author{Robert Presl}
\author{Pabitro Ray}
\author{Andreas Wieser}

\affiliation[ETH Baug]
{Department of Civil, Environmental and Geomatic Engineering, Institute of Geodesy and Photogrammetry, ETH Zurich, Switzerland}

\author{Ursula Keller}
\author{Christopher R. Phillips}

\affiliation[ETH Physics]
{\onehalfspacing Department of Physics, Institute for Quantum Electronics, ETH Zurich, Switzerland}

\title[]{\singlespacing
Long-range and dead-zone free dual-comb ranging for the interferometric tracking of moving targets}

\abbreviations{}
\keywords{dual-comb, ranging, long-distance, dead-zone, real-time, precision}

\begin{document}
\singlespacing

\begin{abstract}
Dual-comb ranging has emerged as an effective technology for long-distance metrology, providing absolute distance measurements with high speed, precision, and accuracy. Here, we demonstrate a dual-comb ranging method that utilizes a free-space transceiver unit, enabling dead-zone-free measurements and simultaneous ranging with interchanged comb roles to allow for long-distance measurements even when the target is moving. It includes a GPU-accelerated algorithm for real-time signal processing and a free-running single-cavity solid-state dual-comb laser with a carrier wavelength \mbox{$\lambda_c \approx$ 1055 nm}, a pulse repetition rate of 1 GHz and a repetition rate difference of 5.06~kHz. This combination offers a fast update rate and sufficient signal strength to reach a single-shot time-of-flight precision of around 0.1~\unit{\micro \meter} (i.e. $< \lambda_c/4$) on a cooperative target placed at a distance of more than 40~\unit{\meter}. The free-running laser is sufficiently stable to use the phase information for interferometric distance measurements, which improves the single-shot precision to $<20~\unit{\nano \meter}$. To assess the ranging accuracy, we track the motion of the cooperative target when moved over 40~\unit{\meter} and compare it to a reference interferometer. The residuals between the two measurements are below 3~\unit{\micro \meter}. These results highlight the potential of this approach for accurate and dead-zone-free long-distance ranging, supporting real-time tracking with nm-level precision.
\end{abstract}

\section{1 Introduction}
Long-range absolute distance measurements are crucial for a wide range of application fields, e.g., satellite navigation and formation flying \cite{Singla2006, Grzegorz2020}, surveying, geodesy and mapping \cite{Müller2008, Joffray2016, Shan2018}, autonomous driving \cite{Royo2019, Li2020, Bastos2021} and the precision manufacturing industry \cite{Gao2015, Schmitt2016, Fengzhou2018}. The demanding requirements of these applications in terms of speed, range, precision, and accuracy propelled the development of various techniques for absolute distance measurements such as pulsed time-of-flight (ToF) LiDAR \cite{Maatta1993, Amann2001}, amplitude-modulated continuous-wave (AMCW) LiDAR \cite{Shan2018, Rüeger1990}, frequency-modulated continuous-wave (FMCW) LiDAR \cite{Culshaw1985, Poulton2017}, and synthetic wavelength interferometry \cite{Dandliker1988, deGroot1991}.

Femtosecond pulsed lasers and optical frequency combs \cite{Telle1999, Apolonski2000, Jones2000, Fortier2019} enable boosting the performance of conventional ranging technologies \cite{Minoshima2000, Matsumoto2003, Lee2010, Baumann2014} and the development of new techniques such as dual-comb ranging \cite{Coddingtion2009, Liu2011, Zhang2014, Chang2024}. Dual-comb ranging is particularly interesting for high-precision applications at long range, as it can combine absolute distance information (from a ToF-based measurement) with the high precision of interferometric measurements. This technique is based on linear optical sampling of a signal pulse train (which probes the distance between a reference and target plane) with a local oscillator pulse train \cite{Coddingtion2009}. This sampling is enabled by the pulse repetition rate difference $\Delta f_\mathrm{rep}$ between the combs. It leads to a down-conversion of the optical time- and frequency domain information to the more accessible electronic domain which allows for precise measurements of the distance between reference and target.

With a source that exhibits a high pulse repetition rate $f_\mathrm{rep}$, the dual-comb ranging approach also offers a fast update rate (determined by $\Delta f_\mathrm{rep}$) which is limited by the aliasing condition to $\Delta f_\mathrm{rep} < f_\mathrm{rep}^2 / (2 \Delta\nu_\mathrm{opt})$ for a given $\Delta\nu_\mathrm{opt}$ (full optical bandwidth shared by the two combs). For example, a pair of \mbox{100-GHz} silicon-nitrite micro-ring resonators enabled an ultra-high update rate of 96~\unit{\mega \hertz} \cite{Trocha2018}. However, the high repetition rate leads to a reduced non-ambiguity range (NAR) in distance measurements. The NAR is given by $R_\mathrm{A, i} = \upsilon_g /(2 f_\mathrm{rep,i})$, where the index $i = 1,2$ indicates the signal comb used to sample the distance between the reference and target plane, and $\upsilon_g$ is the group velocity. For a 100-GHz dual-comb system, $R_\mathrm{A, i}$ is around 1.5~mm in ambient air.


Established dual-comb generation platforms based on femtosecond solid-state and fiber lasers with repetition rates of around 100 MHz offer a much larger NAR. However, such lasers do not support a high update rate unless with a considerably narrowed optical bandwidth \cite{Mitchell2021} or alternative detection schemes \cite{Wright2021}, leading to reduced single-shot precision. This highlights the challenge of combining long-distance measurements with high precision and fast update rates to track moving objects. 


To address this, we utilized a low-noise 1-GHz solid-state dual-comb with a repetition rate difference of $\Delta f_\mathrm{rep} = 5.06~\unit{\kilo \hertz}$ as this provides a reasonably fast update rate for tracking moving targets, while maintaining a broad bandwidth and large photon number per pulse to enable a high single-shot measurement precision. This laser source is combined with a free-space transceiver unit which avoids problematic spurious ghost pulses, enables dead-zone free measurements at all distances by creating two slightly delayed reference reflections, and allows to measure the distance simultaneously with the role of the two combs interchanged to extend the non-ambiguity range using the Vernier effect \cite{Coddingtion2009}, even for targets in motion \cite{Camenzind2022}.

To leverage the high measurement update rate, we developed a GPU-accelerated data acquisition platform to process the acquired electronic signals, extract the ToF distance, and perform the phase-based interferometric ranging in real-time to enable continuous tracking of the target's position with interferometric precision. For this application, the GPU offers faster processing speeds compared to a CPU, particularly for computations such as the Fast Fourier Transform (FFT) which are well-suited for parallelization. Furthermore, GPU based approaches are appealing for dual-comb data processing since they combine real-time processing capabilities with a convenient programming interface \cite{Bresler2023, Walsh2024}. 

The dual-comb ranging setup, including the data-analysis routine, is discussed in Section~2. To assess the accuracy and precision of the dual-comb ranging system, we track the motion of a movable trolley on a linear comparator bench over a distance of 40~\unit{\meter} which is discussed in Section~3. Section~4 summarizes the key results of this work.

\section{2 Experimental setup}
\label{sec:principle_and_design}

\subsection{2.1 Theoretical background on ToF precision}
The precision of dual-comb interferogram (IGM) arrival time estimation can be formulated in terms of a more general ToF estimation framework as follows: The Cramer-Rao Lower Bound (CRLB) for the arrival time of a pulse in a ranging measurement subject to additive white noise is summarized in Ref.~\citenum{Kay1993}. Adapting this bound for the case of dual-comb ranging under the assumption that the measurement is shot-noise limited and that the local oscillator has a much higher average power than the returning signal comb, we find a simple approximate lower bound on the time-of-flight variance due to the additive noise floor:
\begin{align} \label{eq:CRLB}
\sigma^2(\tau_{\mathrm{opt}}) > 
O(1)\cdot \frac{1}{\Delta\nu_{\mathrm{opt}} \cdot f_{\mathrm{rep}} \cdot N_{\mathrm{photon}} },
\end{align}    
where $N_{\mathrm{photon}}$ is the effective number of detected photons of the returning signal beam, and $O(1)$ represents a pulse-shape dependent prefactor. It has been shown that direct ToF calculations can readily be used to obtain \unit{\micro \meter}-level precision \cite{Coddingtion2009, Liu2011, Zhang2014}. Obtaining precise ToF estimates is particularly crucial to leverage the phase information in the IGMs. If the ToF can be reliably determined to better than $\mathrm{\lambda_c}/4$, the distance estimate can be refined by the much more sensitive IGM phase measurements \cite{Coddingtion2009}.

Eq.~\ref{eq:CRLB} shows how gigahertz repetition rate lasers with broad optical bandwidth are preferable in terms of ToF precision. However, another relevant consideration is the NAR. As discussed in Ref.~\citenum{Coddingtion2009}, when the target is measured simultaneously with the role of the combs interchanged, it is possible to infer the number of non-ambiguity ranges to the target without recourse to a separate measurement. To accomplish this, it is necessary to determine the ToF to within one half of the optical delay step between subsequent pairs of pulses which is given by:
\begin{align}
T_{\mathrm{step}}=\frac{\Delta f_{\mathrm{rep}}}{f_{\mathrm{rep}}^2}.
\end{align}
Few-gigahertz lasers represent a sweet spot where a realistic ToF precision is sufficient for both interferometric hand-over and non-ambiguity range unwrapping, while also offering fast (kilohertz) update rates. For example, a 1-GHz dual-comb with 3-THz optical bandwidth, a repetition rate difference of $\Delta f_\mathrm{rep} = 5~\unit{\kilo \hertz}$ and a wavelength of 1~\unit{\micro \meter} requires a ToF estimation to within 2.5~\unit{\femto\second} for non-ambiguity range estimation and 1.7~\unit{\femto\second} for interferometric hand-over. Applying these values to Eq.~\ref{eq:CRLB} assuming a 1-\unit{\micro\watt} returning power and a photodiode responsivity of 0.7~\unit{\ampere/\watt} yields a CRLB of 0.6~\unit{\femto\second}, i.e. sufficient for both steps assuming peak finding at the CRLB limit.

Based on the above considerations, gigahertz solid-state lasers are compelling sources for precise and long-distance dual-comb ranging as they offer kilohertz measurement update rates, multi-THz bandwidths, low-noise properties, and a direct free-running oscillator-based setup with minimal complexity and substantial average power. We demonstrate this potential via a novel dual-comb laser source, a new transceiver unit design, and a real-time data processing routine.


\subsection{2.2 Laser Source}

\begin{figure}[!htbp]
    \centering
    \includegraphics[scale = 1.0]{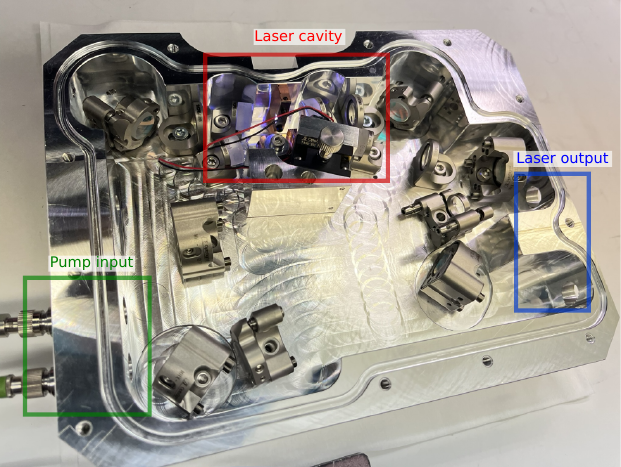}
    \caption{Picture of the laser in a prototyped housing. The pump input is highlighted in green, the laser cavity in red and the laser output in blue. Not shown in this image are the fiber collimation units which were later installed in the laser output to couple the light into single-mode fibers.}
    \label{fig:laser}
\end{figure}

The laser source is a gigahertz single-cavity dual-comb laser conceptually similar to the one presented in Ref.~\citenum{Willenberg2024}. The laser is shown in Fig.~\ref{fig:laser} with open lid. It uses single-mode diode pumping and a spatially multiplexed cavity \cite{Pupeikis2022}. Compared to the Yb:CALGO laser in \mbox{Ref.~\citenum{Willenberg2024}}, the main modifications are (i) the use of a different gain medium, Yb:CaF$_2$; (ii) use of two pump diodes instead of one; and (iii) integration of the laser into a robust prototype setup. The Yb:CaF$_2$ gain material is beneficial in the context of low-noise lasers since it has a lower nonlinear refractive index than Yb:CALGO \cite{Sevillano2013, Boudeile2007}, and offers a small emission cross section which enables a low relaxation oscillation frequency and therefore strong low-pass filtering of pump intensity noise \cite{Seidel2023}.

The laser operates in the soliton modelocking regime \cite{Kartner1996} with a nominal round-trip group delay dispersion (GDD) of -150~\unit{\femto\second^2}, and modelocking is obtained by a semiconductor saturable absorber mirror (SESAM) \cite{Keller1996} with $\approx 1~\unit{\pico\second}$ recovery time. The combs are coupled into individual single-mode fibers for beam delivery to the transceiver unit with about 70\% efficiency. The performance characteristics are summarized in Table \ref{tab:laser}.

\begin{table}
\centering
\begin{tabular}{ |l|c|c c| } 
 \hline
 \phantom{x} & Unit & Comb 1 & Comb 2 \\ 
 \hline 
 Center wavelength                   & \unit{\nano\meter} & 1056 & 1055 \\
 $P_{\mathrm{average}}$ (oscillator) & \unit{\milli\watt} & 108 & 100 \\ 
 $P_{\mathrm{average}}$ (fiber)      & \unit{\milli\watt} & 75 & 70 \\ 
 Bandwidth (oscillator)              & \unit{\nano\meter} & 14 & 10 \\
 Bandwidth (fiber)                   & \unit{\nano\meter} & 22 & 17 \\
 $f_{\mathrm{rep}}$ (nominal)       & \unit{\giga\hertz} & \multicolumn{2}{c|}{1.041}  \\
 $\Delta f_{\mathrm{rep}}$ (tuning range)  & \unit{\kilo\hertz} & \multicolumn{2}{c|}{0 to 200} \\
 \hline
\end{tabular}
\caption{Dual-comb laser parameters. The average power $P_{\mathrm{average}}$ is specified for both the direct oscillator output and after the beam delivery fibers. Bandwidth refers to the full width at half maximum (FWHM).} 
\label{tab:laser}
\end{table}

Intensity noise of the pump laser leads to timing jitter on the two combs. This issue was examined in detail for similar lasers before \cite{Seidel2023, Willenberg2024}. In Ref.~\citenum{Willenberg2024}, correlated timing jitter fluctuations were obtained by splitting a single pump diode into two equal parts to pump each comb. Here, we adopt a different approach by using two pump diodes combined through a 50:50 fiber splitter so that each combs' pump contains around half of the power from each diode. This results in correlated pump power fluctuations, hence reducing the pump power induced effects on the timing jitter of $\Delta f_{\mathrm{rep}}$. 




The two low-noise combs emitted by the laser are coupled into FC/APC single-mode fibers inside the prototype housing to deliver the light to the transceiver unit which will be discussed next.

\subsection{2.3 Dual-comb transceiver unit}

Our previous dual-comb ranging setup used a fiber-based transceiver unit \cite{Camenzind2022}, which enabled alignment-free operation while measuring the target distance with both combs simultaneously acting as signal comb and each others local oscillator, thereby enabling Vernier-based non-ambiguity range unwrapping \cite{Coddingtion2009}. An FC/PC output fiber was used to generate a reference reflection from its end face. However, we found that the fiber-based setup exhibited multiple etalon-like effects leading to ghost pulses that can reduce the measurement precision when they are close to the target reflection. These effects arose from polarization projections due to imperfections in the fiber components, and from a ``cavity'' formed between the output fiber tip and the target. Additionally, the reference reflection also led to a so-called dead-zone \cite{Lee2013} when the target is an integer number of cavity lengths away from the fiber tip, as this causes the target and reference reflections to arrive at the same time and therefore interfere with each other. Such ghost-pulses and dead-zones can limit the applicability of a ranging system for continuous measurements. 

\begin{figure}[!htbp]
    \centering
    \includegraphics[scale = 1.1]{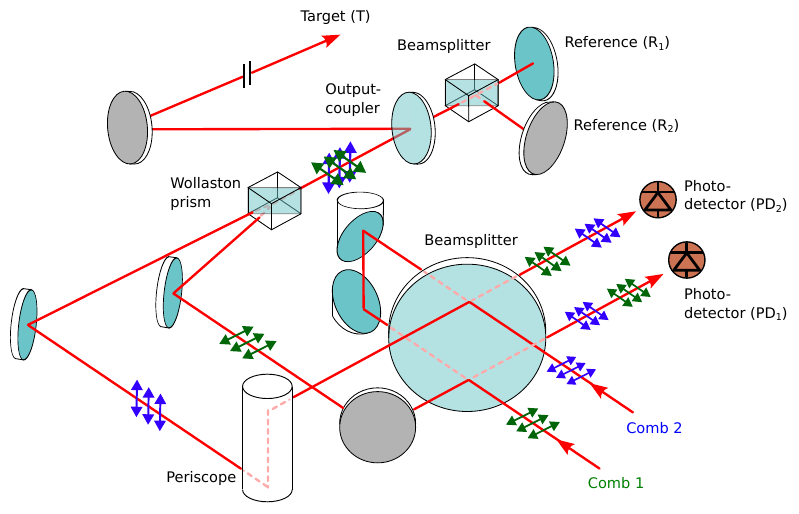}
    \caption{Schematic of the dual-comb transceiver unit with the polarization of comb 1 (comb 2) indicated in green (blue). The input polarization was adjusted with fiber polarization controllers on the delivery fiber from the laser source to the transceiver unit.}
    \label{fig:frontend}
\end{figure}

To address these issues, we have developed a new setup that offers dead-zone free measurements at any distance. Fig.~\ref{fig:frontend} shows a schematic of this new transceiver unit. Similar to the implementation in Ref.~\citenum{Camenzind2022}, it also allows measuring the target distance simultaneously with the roles of the combs interchanged. In particular, we are combining the orthogonally-polarized combs with a Wollaston prism and send them together through a polarization-maintaining (PM) single-mode fiber for mode-matching so that they are colinear along the measurement path. This ensures that both combs travel the same distance along the propagation path between the reference and target. On $\mathrm{PD}_i$, ($i = 1,2$) we can then record the distance probed with comb $i$, while comb $j$, ($j = 2,1$) serves as local oscillator to map the optical signal pulses reflected by both references and target into the electronic time domain. This downconversion allows for easy detection of these signals with a photodiode (DET08C/M, Thorlabs) and subsequent electronic amplification (ENA220-T, RF-Bay), while still maintaining the high measurement precision inherent to the optical time domain. We use the two measurements to deduce the number $m \in \mathbb{N}_0$ of non-ambiguity ranges $R_\mathrm{A, i}$ to the target, thereby extending the effective non-ambiguity range to several kilometers using the Vernier effect. Furthermore, since these measurements are recorded simultaneously, we can perform this computation for each $\Delta f_{\mathrm{rep}}$ period of the data which allows tracking moving targets at long distances \cite{Camenzind2022}. 

\begin{figure}[!htbp]
    \centering
    \includegraphics[scale = 1.1]{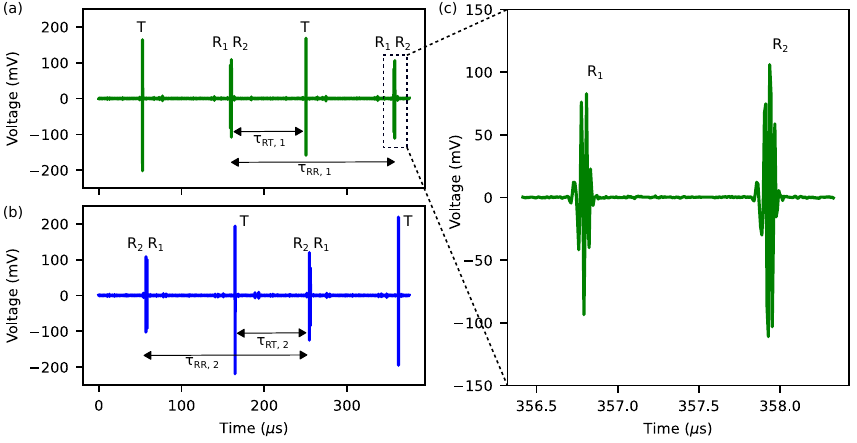}
    \caption{(a) Interferogram trace which shows the reference reflections ($R_1$ and $R_2$) and target reflection (T). The delay $\tau_{\mathrm{RT, 1}}$ encodes the separation between reference and target, while the delay $\tau_{\mathrm{RR, 1}} = 1/\Delta f_{\mathrm{rep}} = 198~\unit{\micro \second}$ indicates the measurement update rate. (b) Interferogram signal recorded on the second photodetector with the roles of signal beam and local oscillator interchanged. This simulteaneously-recored second channel allows to extract the number of non-ambiguity ranges on each measurement period. (c) Zoom on the two reference IGMs which allow for dead-zone free measurements.}
    \label{fig:IGM}
\end{figure}

The transceiver unit mainly relies on free-space components to suppress unwanted reflections or polarization projections that lead to etalon-like features in the IGMs. Although such reflections are still present in our measurement, they are close to the noise floor as visible in Fig.~\ref{fig:IGM}(a, b) which shows typical interferogram traces. The IGM signals correspond to the light reflected by the target (T) and the two references ($R_1$ and $R_2$). The reference reflections are generated outside the target path using an output coupler (see Fig.~\ref{fig:frontend}), as this prevents the formation of a cavity between the reference and target planes. The delay $\tau_\mathrm{RR, i} = 1/\Delta f_\mathrm{rep}$ between consecutive pairs of signals from the same reference (i.e. R$_1$ to R$_1$ and R$_2$ to R$_2$) provides the repetition rate difference and the delay $\tau_\mathrm{RT, i}$ between the target and reference reflections encodes the measurement distance. Thus, the absolute distance based on the time-of-flight information is
\begin{equation}
    d_\mathrm{ToF} = \frac{\upsilon_g}{2} \cdot \tau_{\mathrm{RT}} + m \cdot R_\mathrm{A, i}.
\end{equation}

To avoid dead-zones from the overlap of reference and target interferograms,  the transceiver unit creates two reference reflections with a small offset relative to each other as visible in Fig.~\ref{fig:IGM}(c). This ensures that there is always at least one reference not overlapping with the target IGM. In case the target IGM would overlap with one of the reference IGMs, i.e. if we are within the dead-zone, the second (undistorted) reference can be still be used to (i) get an accurate estimate of the reference position; (ii) subtract the distorted reference IGM to clean the target IGM and thereby enable dead-zone free measurements as discussed in detail in the next section; and (iii) infer and correct for the laser timing fluctuations by tracking $\Delta f_\mathrm{rep}$ via $\tau_\mathrm{RR, i}$. To also account for fluctuations in $f_\mathrm{rep}$, we record the down-mixed product between the pulse repetition rates and a 1-GHz output of a signal generator, alongside the interferometric dual-comb signals, on our data acquisition card (M4i.4451-x8, Spectrum Instrumentation). The clock of the data acquisition card and the signal generator are both synchronized to the same 10-\unit{\mega \hertz} Rubidium atomic clock (FS725, Standford Research Systems) which enables tracking the absolute value of $f_\mathrm{rep}$.

 
 
 

\subsection{2.4 Data analysis}


Real-time phase- and timing correction of the dual-comb interferograms is a critical aspect for obtaining large datasets, long integration times, or deploying such systems to real-world measurements. Two of the most popular strategies are solutions based on FPGAs \cite{Roy2012} and GPUs \cite{Bresler2023, Walsh2024}. FPGAs are advantageous in terms of low power consumption and low latency, but GPUs are more user-friendly as they can be programmed with standard programming interfaces such as CUDA (as opposed to the hardware description tools necessary for FPGA development). There are several data acquisition cards with established interfaces to the GPU, making this a generally accessible approach. Here we use a GPU approach based on the M4i.4451-x8 digitizer card (Spectrum Instrumentation), operating at a sampling rate of 125 MS/s. Our approach is conceptually similar to those discussed elsewhere, but adapted to the particular data processing steps involved with the new transceiver unit.


\begin{figure}[!htbp]
    \centering
    \includegraphics[scale = 0.8]{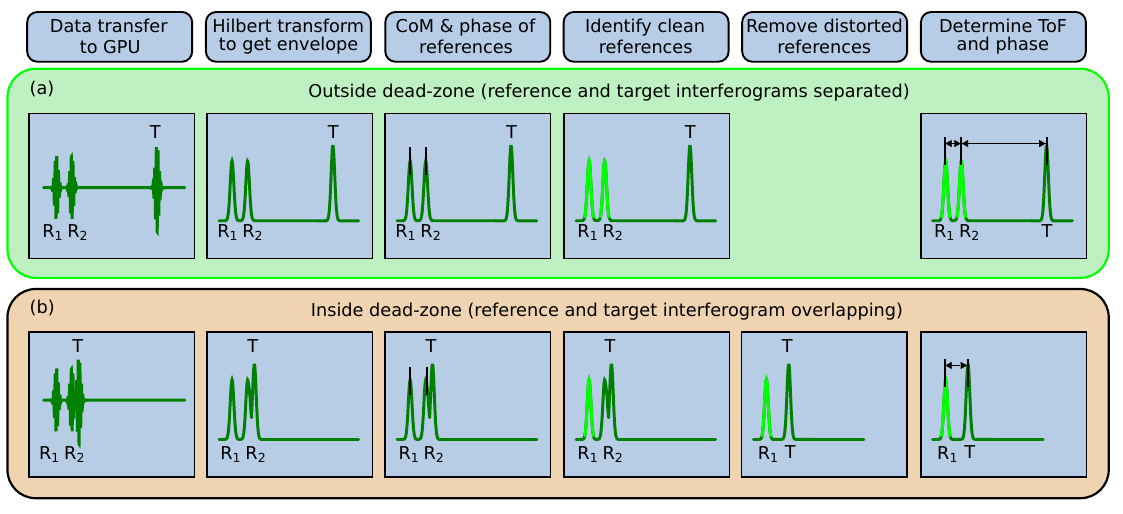}
    \caption{Relevant processing steps of the GPU-accelarated algorithm for the time-of-flight and phase-based distance measurement (a) outside the dead-zone and (b) within the dead-zone. The clean references, i.e. those not overlapping with the target IGM, are indicated in light green. T: target interferogram, R$_i$: reference interferograms, CoM: center-of-mass.}
    \label{fig:GPU}
\end{figure}

An overview of the data processing steps performed by the real-time GPU algorithm is shown in Fig.~\ref{fig:GPU}. We distinguish between the cases outside the dead-zone, i.e. when the reference and target IGMs do not overlap (Fig.~\ref{fig:GPU}(a)), and within the dead-zone, i.e. when the target IGM is distorted (Fig.~\ref{fig:GPU}(b)). In both cases, the data is first transferred to the GPU, followed by a frequency-shift of the dual-comb signal towards DC and a Hilbert transform to extract the envelope. We then compute (i) the phase delay between the two references $R_1$ and $R_2$, (ii) the phase delay to the corresponding references in the previous period and (iii) the center-of-mass of the references using a second-order moment integral where we additionally filter out signals below a certain threshold to be less susceptible to ghost pulses.

To allow for a comparison between references, we then use this information to compute the difference between the reference IGMs by subtracting one interferogram from the other while accounting for the measured difference in their amplitude and phase. By comparing the two references $R_1$ and $R_2$ with each other and to the corresponding reference in the previous period, we can determine if one of them overlaps with the target, and also identify which reference overlaps if there is one. The references not overlapping with the target IGM are marked as clean references.

\begin{figure}[!htbp]
    \centering
    \includegraphics[scale = 1.1]{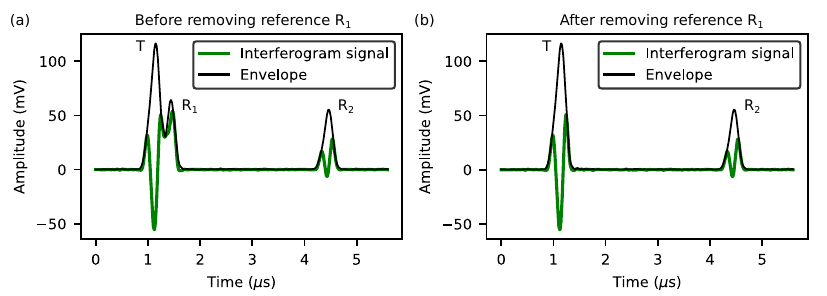}
    \caption{(a) Example trace showing the situation of overlapping target and reference IGMs (here $\mathrm{R_1}$ is overlapping with T), which was extracted during the processing of the actual measurement data. (b) By subtracting reference $\mathrm{R_2}$ from $\mathrm{R_1}$, while accounting for the measured difference in their position, amplitude, and phase, we suppress $\mathrm{R_1}$ and thereby clean the target IGM to avoid systematic errors in the center-of-mass estimation caused by their overlap.}
    \label{fig:Remove Reference}
\end{figure}

The center-of-mass and phase-delay computed for the clean references allows to update the measurement parameters such as e.g., $\Delta f_{\mathrm{rep}}$, and clean the target IGM in case it overlaps with a reference by subtracting the clean reference from the distorted reference as illustrated in Fig.~\ref{fig:Remove Reference}. For this subtraction procedure, we account for the measured difference in the position, amplitude, and phase between the references (based on the most recent measurement where both references were clean). Finally, we obtain clean reference and target IGMs from which we can calculate the time-of-flight and phase delay required for extracting the distance information. To make the Vernier-based non-ambiguity range extension more robust against small fluctuations in the measured ToF-distance, we calculate the number $m$ of non-ambiguity ranges with an exponential moving average and restrict changes to $\pm~1$ from one period to the next.



It is important that the algorithm achieves a sufficiently high throughput: if it is too low, then $\Delta f_{\mathrm{rep}}$ must be reduced accordingly to allow for real-time processing, and eventually this precludes fully coherent measurements from a free-running dual-comb laser. In our case, the full algorithm (including both ToF and phase extraction) could be implemented at up to around 7.1~\unit{\kilo\hertz}. A limiting factor of the processing time is the Hilbert transform, which requires the computation of two FFTs. In our case, this step alone can take up to $\approx$ 100~\unit{\micro \second} on our GPU (RTX A5000, NVIDIA) when working with the single-precision floating point format (FP32). On another PC equipped with an Intel Core i7-9700K CPU and 32~GB RAM, the Hilbert transform takes a longer processing time of around 1~\unit{\milli \second} which shows the advantages of GPU based calculations. To further keep the processing time of each period consistent and as short as possible, the algorithm is designed to only use element-wise and reduction operations such as summations or maximum finding, rather than relying on iterative processes such as curve fitting. 





Because the new laser employs a robust prototype cavity construction (see Fig.~\ref{fig:laser}), it is compatible with coherent averaging down to $\Delta f_{\mathrm{rep}}$ below 1~\unit{\kilo\hertz}. This enables fully coherent data processing with the GPU-accelerated algorithm. To operate between these limitations imposed by coherent averaging capability and GPU-processing speed, and to also balance update rate with the effective number of detected signal photons per period $N_\mathrm{photon} \propto 1/\Delta f_\mathrm{rep}$ which is beneficial for the measurement precision (see Eq.~\ref{eq:CRLB}), we chose a repetition rate difference of $\Delta f_\mathrm{rep} = 5.06~\unit{\kilo\hertz}$. These considerations also highlight that, when targeting high update rate measurements with free-running dual combs, it is important to consider both the constraints imposed by the laser and those related to the data processing. For the data analysis presented in the following section, we recorded several long time traces of $\approx 500~\unit{\second}$ and used them as input for our real-time processing code. This allowed us to refine and optimize our algorithm "offline" using one of the traces before ultimately applying the algorithm to process another trace to obtain the data shown in this manuscript. Here we confirmed real-time processing capability by ensuring a processing rate faster than $\Delta f_\mathrm{rep}$: the processing of each period takes $\approx 140~\unit{\micro \second}$ corresponding to the aforementioned maximum update rate of 7.1~\unit{\kilo \hertz}.

\section{3 Results}
\label{sec:results}

\subsection{3.1 Measurement bench}

\begin{figure}[!htbp]
    \centering
    \includegraphics[scale = 1.1]{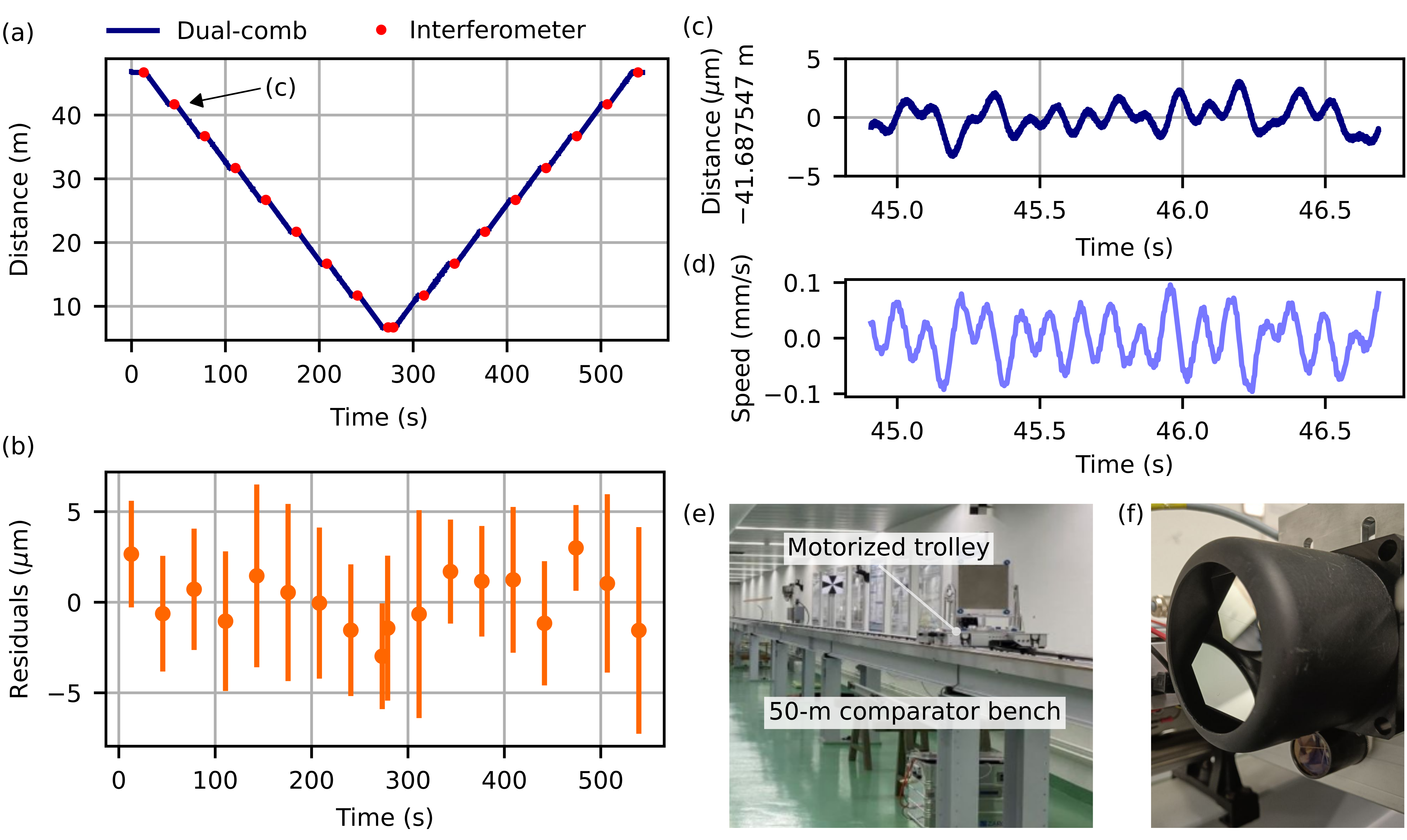}
    \caption{(a) Measurements obtained with the dual-comb ranging setup (using the time-of-flight information), compared with reference measurements from a Doppler interferometer. (b) Residuals of the dual-comb measurement with respect to the interferometric measurement: mean (dot) and maximum variation (error-bars) represent the slow oscillations across the static regions. (c) Measured distance during the static region at around 45~\unit{\second} which exhibits slow oscillations and (d) corresponding velocity. (e) Image of the 50-m comparator bench used for the long-distance measurements and (f) the two retroreflectors used for the dual-comb ranging (top) and reference interferometer (bottom), respectively.}
    \label{fig:Accuracy}
\end{figure}

We conducted experiments in controlled laboratory conditions on a 50-\unit{\meter} long linear comparator bench (see Fig.~\ref{fig:Accuracy}(e)). The comparator bench is equipped with a computer-controlled motorized trolley that can move along its entire length. The bench further includes a Doppler interferometer based on a Helium-Neon laser (HP5519A, Keysight) to take calibrated snapshots of the distance. Our dual-comb apparatus was positioned next to the bench and we directed the combs emitted by the transceiver unit to an optical breadboard mounted on the bench to launch them towards the target. On the breadboard, the beams are first expanded and then collimated with a pair of lenses forming a Keplerian telescope to prevent significant beam divergence along the measurement path. This telescope is also used to collect the back-reflected light. The combs were aligned parallel to the reference interferometer beam along the measurement path. The target is a retroreflector mounted on the movable trolley. There is also a second retroreflector (mounted below the target retroreflector) for the reference measurement with the interferometer (see Fig.~\ref{fig:Accuracy}(f)). 

Since there is a non-zero baseline between these retroreflectors ($\approx 5.6~\unit{\centi \meter}$), small tilts of the trolley caused e.g. by bending of the rails introduces an offset between the distance measured with the two systems. This offset is known as Abbé-error. To estimate this offset, we measure the inclination of the trolley with an inclination sensor (065-040TYPE3-10, Zerotronic) as it moves along the comparator bench. Together with the baseline between the retroreflectors, we then approximate the Abbé-error and correct it in the reference measurement.

Our objectives for the measurement campaign included demonstrating real-time measurements, maintaining absolute calibration over long distances (here demonstrated over a 40 m range), conducting continuous measurements for several minutes, and enhancing the precision by exploiting the IGM phase information for interferometric measurements. To show these capabilities, the trolley was initially positioned at around 50 m and was automatically moved closer in 5-m steps to around 10 m, remaining stationary at each intermediate position for 5~\unit{\second}, as depicted in Fig.~\ref{fig:Accuracy}(a). At each step, we recorded a reference measurement of the distance with a Doppler interferometer. Similarly, the trolley was then moved again to around 50~m in 5~m increments to evaluate any systematic deviations relative to the reference measurements and to assess the temporal stability of the experimental setup. The software controlling the stop-and-go movement of the trolley outputs a single interferometer distance reading after the trolley stops at each location, with time stamps accurate to within few 0.1~s. We therefore only used the interferometer readings to check for systematic offsets in the absolute distance measurement at the stationary trolley positions. The reference measurement accounts for the phase refractive index in air at approximately 633~nm (Helium-Neon laser). We also scale the ToF-based dual-comb ranging measurement with the inverse of the group refractive index $n_\mathrm{g}$ at $\lambda_c$ calculated from the corrected Ciddor's equation provided in \mbox{Ref.~\citenum{Pollinger2020}} using the meteorological parameters continuously monitored in the lab. 

We launch around 1~\unit{\milli\watt} of optical power per comb towards the retroreflector target. The distance to the target affects the coupling efficiency of the back-reflected light into the PM single-mode fiber and thus the power on the photodetectors. Typically, it is tens of \unit{\micro \watt} per comb. 

\subsection{3.2 Accuracy}
For each stationary position of the trolley, we compare the mean of the ToF-based dual-comb ranging measurements during this static region to the reference measurement obtained from the Doppler interferometer. From Fig.~\ref{fig:Accuracy}(b) we find that the residuals are below 3~\unit{\micro \meter} across the 40~\unit{\meter} range. The error bars in Fig.~\ref{fig:Accuracy}(b) represent the variation of our dual-comb ToF-based distance measurements across the different static windows at each position. They are dominated by a slow oscillation caused by external instabilities resulting in real or apparent distance changes with amplitudes on the order of a few \unit{\micro \meter} and a dominant frequency of around 10 Hz. This is illustrated in Fig.~\ref{fig:Accuracy}(c) for the measurement at the static position where the trolley arrives after around 45~\unit{\second}. Fig.~\ref{fig:Accuracy}(d) shows the corresponding velocity determined from the ToF data after applying a low-pass filter to extract the slow oscillation.

Without e.g. tracking the velocity, the non-ambiguity range extension is only possible up to a delay of $\Delta \tau = 0.5 \cdot \Delta f_{\mathrm{rep}}/f_{\mathrm{rep}}^2$ from one period to the next. In our case, this corresponds to a maximum target speed of roughly 1.77~\unit{\milli \meter}/\unit{\second}. As apparent from Fig.~\ref{fig:Accuracy}(d), it would thus not be necessary to track the target speed within the static regions. However, to also enable accurate measurements when the trolley is moving between the static regions, our system accounts for the target's speed by interpolating the distance measurements to a common reference time between the channels \cite{Camenzind2022}. When tracking and correcting for velocity, a more general limit exists also with respect to acceleration. 



\subsection{3.3 Precision}

Next, we consider the precision of the ToF-based distance measurement. Since the available reference measurements undersample the motion of the target, we need an alternative means to infer the precision of the dual-comb ranging measurement. For that purpose, we consider short time windows of 5 periods ($= 5/\Delta f_\mathrm{rep} \approx 1~\unit{\milli \second}$), compute a linear fit through the measurements within each time window, and calculate the standard deviation of the resulting residuals compared to the fit to quantify the precision. To furthermore account for the dependence of precision on the relative delay of reference and target interferograms \cite{Shi2015}, these time windows are from measurements where the target is moving between two static regions. 

\begin{figure}[!htbp]
    \centering
    \includegraphics[scale = 1.1]{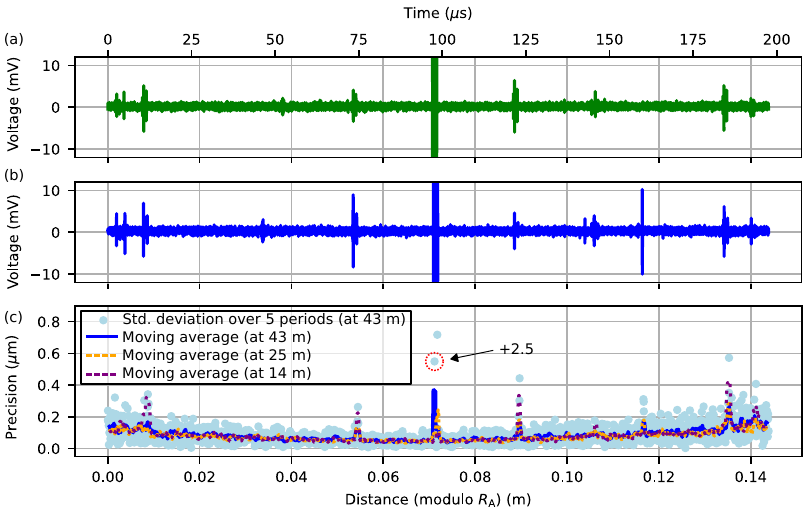}
    \caption{Vertical zoom on one period of the IGM trace recorded with (a) the first photodetector ($\mathrm{PD_1}$) and (b) the second photodetector ($\mathrm{PD_2}$) with inverted time-axis. These example traces were selected and aligned such that the target (T) and reference (R) IGMs are both located at the center of the window. (c) Ranging precision for different offsets between T and R, calculated from the standard deviation of 5 distance measurements w.r.t their linear fit for measurements at around 43~m where the target is moving so that the delay between T and R changes. This delay $\tau_\mathrm{RT}$ is encoded in the distance $\upsilon_g/2 \cdot \tau_\mathrm{RT}$ (modulo the non-ambiguity range $R_\mathrm{A}$). The highlighted point has been shifted down by 2.5~\unit{\micro\meter} to keep the rest of the features visible. We also compute a moving average (window size of 10 points) over the measurements around 43~m (blue), 25~m (orange) and 14~m (purple).}
    \label{fig:Precision}
\end{figure}

More specifically, since the target is moving, the precision measured for the individual time windows corresponds to different absolute distances and thus different delays between target and reference IGMs. Generally, shorter delays between reference and target lead to reduced relative timing jitter and hence improved precision \cite{Shi2015}. To account for this, we sort the measurements according to the absolute distance modulo the non-ambiguity range $R_\mathrm{A, i}$. This analysis routine was performed for regions around 43~m, 25~m, and 14~m which all indicate similar precision as shown in Fig.~\ref{fig:Precision}(c). The center of the plot corresponds to the measurements when the target IGM overlaps with the reference IGMs, and the edges of Fig.~\ref{fig:Precision}(c) correspond to the situation with the target IGM furthest away from the reference IGMs on the acquired electronic signal. For this latter case, the relative timing jitter between reference and target is highest \cite{Shi2015}, which explains the degradation of precision towards the edges in Fig.~\ref{fig:Precision}(c). For most of the measurements, the point-to-point fluctuations have a standard deviation of $<0.1~\unit{\micro \meter}$. At a few specific positions, the precision degrades from $\ll 1~\unit{\micro \meter}$ to $\approx 1~\unit{\micro \meter}$. 

In Fig.~\ref{fig:Precision}(c), the degradation in the center originates from the reference IGMs. Having a pair of reference reflections helps to subtract the reference part of the waveform when it overlaps with the target reflection (see Fig.~\ref{fig:Remove Reference}). However, since this subtraction is not always perfect and adds noise of the subtracted signal to the resulting signal, the precision for those points may be reduced. Alternative approaches to circumvent this issue are to separate reference and target pulses based on their polarization \cite{Lee2013, Han2024}, or to have the references on a separate path, i.e. to not have them propagate back with the target reflection. While these approaches avoid the issue of overlapping IGMs, they require additional detectors and data channels making the electronics more complex. Having the references on a separate path might furthermore lead to reduced accuracy due to systematic offsets between the reference and signal paths (since they would no longer have a near-common-path architecture). 

The remaining positions for which the precision degrades in Fig.~\ref{fig:Precision}(c) originate from ghost pulses which are revealed when zooming in on the vertical axis of the dual-comb signals recorded with the two photodetectors. To visualize this, we selected and aligned example traces recorded with the first photodetector (Fig.~\ref{fig:Precision}(a)) and second photodetector (Fig.~\ref{fig:Precision}(b), with inverted time-axis) to the measurements in Fig.~\ref{fig:Precision}(c) such that the reference and target IGMs are in the center. This illustrates that while the free-space setup can significantly reduce ghost-pulses compared to the fiber-based implementation of the transceiver unit \cite{Camenzind2022}, such ghost pulses still remain a limiting factor that causes degradation of the precision at certain measurement distances.

\subsection{3.4 Interferometric hand-over}




Next, we investigate the phase information in the IGMs, which has the potential to significantly improve the measurement precision \cite{Coddingtion2009}. Since our dual-comb IGMs are sufficiently stable for coherent averaging in free-running operation at the selected update rate, phase information can be tracked unambiguously as long as the Doppler shift due to the target's movement does not introduce an interferogram frequency shift of about $\pm \Delta f_{\mathrm{rep}}/2$. More specifically, the dual-comb IGMs encode the instantaneous optical transfer function (OTF) associated with the measurement setup, i.e. the spectral phase associated with free space propagation to the target and back. However, since we are using free-running dual combs with unknown carrier-envelope offset frequency $f_\mathrm{CEO}$ and unknown comb line indices, there is an unknown offset in mapping the measured RF comb line frequencies to the corresponding optical comb line frequencies. While this precludes recovery of the true OTF, the optical frequencies are still known reasonably well due to the known center wavelength measured with an optical spectrum analyzer. Therefore, if we utilize the ToF data to calibrate the absolute optical delay associated with a particular target position, small distance changes around that position can be inferred by the phase instead, thereby improving precision.

To achieve this, we use our GPU-accelarated algorithm to compute and unwrap the phase delay from reference to target IGM alongside the ToF information. Since the resulting phase $\phi(t)$ is proportional to the center wavelength (up to some offset influenced by $f_{\mathrm{CEO}}$), we can employ these phase changes to track relative distance changes with interferometric precision. To get absolute distance measurements with interferometric precision, we can use the ToF information to shift the interferometric distance measurement once by $d_\mathrm{ToF}$ and then continue tracking the distance based on the interferometric phase information, resulting in:
\begin{equation}
    d_\mathrm{phase}(t) = \phi(t) \cdot \frac{\lambda_c}{4 \pi \cdot n} + d_\mathrm{ToF}, 
\end{equation}
where $n$ is the phase refractive index at $\lambda_c$ which can be estimated from empirical equations, e.g. Ciddor's equation \cite{Ciddor1996}. Alternative schemes for the interferometric hand-over rely on counting the numbers of optical wavelengths based on the ToF measurement to achieve interferometric precision combined with absolute distance measurement. While our ToF precision would be sufficient for this approach (ToF precision below $\lambda_c/4$), we lack information about the absolute phase offset due to not tracked and free-running $f_{\mathrm{CEO}}$. 

    

\begin{figure}[!htbp]
    \centering
    \includegraphics[scale = 1.1]{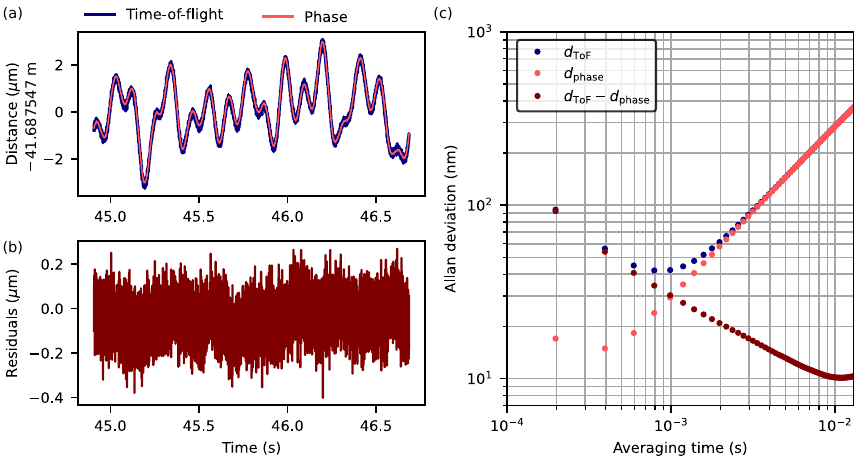}
    \caption{(a) Time-of-flight (blue) and phase-based (pink) distance measurement during the static region at around 45~\unit{\second}. (b) Residuals between the time-of-flight (ToF) and phase-based distance measurements. (c) Allan deviation for the distance measurement based on ToF ($d_\mathrm{ToF}$, blue) and phase ($d_\mathrm{phase}$, pink) information. It also shows the Allan deviation for the residuals in (b), i.e. for $d_\mathrm{ToF} - d_\mathrm{phase}$ which indicates the ToF-precision without the influence of the slow oscillation on a 100-ms timescale. }
    \label{fig:Phase}
\end{figure}

In Fig.~\ref{fig:Phase}(a), we examine the phase-based interferometric measurements overlaid with the ToF measurement for the static-target segment at around 45~\unit{\second} shown previously in Fig.~\ref{fig:Accuracy}(c). The phase-based distance measurements (in pink) are in close agreement with the ToF-based measurements (in blue). The residuals between the two measurements are shown in Fig.~\ref{fig:Phase}(b). 

As apparent from the Allan deviation in Fig.~\ref{fig:Phase}(c), the interferometric hand-over significantly improves the single-shot precision from around 0.1~\unit{\micro \meter} for the ToF-based measurement (blue) down to less than 20~\unit{\nano \meter} for the interferometric measurement (pink). Averaging improves the measurement precision at first, but after an averaging time of around 1~\unit{\milli \second} the precision starts degrading due to the slow oscillation. To compensate for the precision degradation due to this slow oscillation, we can also consider the Allan deviation for the residuals in Fig.~\ref{fig:Phase}(b). Due to the high precision of the interferometric measurement, this can serve as an indicator for the true precision achievable with the ToF-based measurement under suitable conditions (vacuum, stable setup), which reaches around 10~\unit{\nano \meter} after averaging for 10~\unit{\milli \second}.

\section{4 Conclusion}
\label{sec:Conclusion}

We demonstrate a ranging system based on a free-running single-cavity Yb:CaF$_2$ 1-GHz dual-comb laser combined with a free-space transceiver unit optimized to suppress problematic spurious ghost pulses. The system allows for simultaneous distance measurements with the role of the two combs interchanged to track moving targets even at a long range \cite{Camenzind2022}. By generating two reference reflections with a short delay relative to each other, the setup can further circumvent dead-zones caused by regions where the target and reference IGMs overlap. To additionally exploit the fast measurement update rate of 5.06~\unit{\kilo \hertz}, we developed a GPU-accelerated algorithm which enables real-time data processing.


The performance of the dual-comb ranging instrument is tested on a 50-m-long comparator bench equipped with a cooperative target placed on a movable trolley. Comparison to a reference interferometer suggests an accuracy of around $\pm$ 3~\unit{\micro \meter} for repeated measurements every 5~\unit{\meter} over 40~\unit{\meter}. We achieve a single-shot time-of-flight precision of around 0.1~\unit{\micro \meter}, i.e. below $\lambda_c/4$. Due to the laser's low-noise properties (inherent to the prototype housing and low-pass filtering of pump noise), we can track the phase delay between reference and target IGMs even in free-running operation. This allows for interferometric distance tracking which yields a single-shot precision below 20~\unit{\nano \meter}.

The reported accuracy and precision indicate the achievable performance in a controlled laboratory environment, where the meteorological parameters were continuously monitored to calculate the refractive index of air. In practical outdoor conditions, the achievable measurement accuracy is affected by spatial and temporal variations in the refractive index of air along the measurement path, resulting in apparent distance errors. Compensating for these errors requires monitoring meteorological parameters at multiple locations along the measurement path \cite{Pollinger2012} or using simultaneous distance measurements at two or more wavelengths for dispersion-based refractivity correction \cite{Bender1965}. 

Our results showcase the potential of the proposed dead-zone free dual-comb ranging approach for applications requiring real-time and accurate long-distance measurements in controlled environments such as industrial process monitoring, as well as in space applications, including satellite positioning and formation flying. 


\subsection{Funding}
This work was supported by a BRIDGE Discovery Project Nr. 40B2-0\_180933 a joint research programme of the Swiss National Science Foundation (SNSF) and Innosuisse–the Swiss Innovation Agency. This project has received funding from the Swiss National Science Foundation under grant Nr. 200021\_184988.

\subsection*{Disclosures}
LL, BW and JP declare partial employment with K2 Photonics.

\subsection{Data availability}
Data underlying the results presented in this paper is available at ETH Zurich Research Collection library.

\section{Supplementary information}
\subsection{Cramer-Rao lower bound for time-of-flight measurements}

The Cramer-Rao lower bound (CRLB) for a ranging measurement subject to white Gaussian noise is discussed in Ref.~\citenum{Kay1993}. The lower bound on the variance on the inferred timing is given by 
\begin{align}
    \sigma^2_\mathrm{ToF}(T_0) = \frac{\sigma^2_\mathrm{WN} }{2B \int \left(\frac{ds(t)}{dt}\right)^2dt}
    \label{eq:var_tof}
\end{align}
for additive white noise with variance $\sigma^2_\mathrm{WN}$, and a signal $s(t)$ with a full bandwidth $2B$ around which the waveform is filtered to avoid unnecessary noise contributions from other frequencies. 

We can adapt this formula to the case of the positive Fourier frequency branch of a dual-comb interferogram with shot noise limited measurement noise. For simplicity we make the assumption that the local oscillator current $i_\mathrm{LO}$ dominates the signal current $i_\mathrm{signal}$, so that the one-sided shot noise current power spectral density (PSD) is $2qi_\mathrm{LO}$, and that the IGM signal is given by the following expression:
\begin{align}
    \bar{s}(t) =&
    2\sqrt{i_\mathrm{LO} i_\mathrm{signal}} \cdot \mathrm{sinc}\left(
        \frac{t}{T_\mathrm{signal}} \right) \cos(2\pi i \nu t)
        \nonumber \\
        =&
        s(t) 
        \left(e^{2\pi i \nu t} + e^{-2\pi i \nu t}\right),
\end{align}
where $\nu$ is the optical frequency, $T_\mathrm{signal}$ determines the IGM width, and the second expression introduces the signal $s(t)=\sqrt{i_\mathrm{LO} i_\mathrm{signal}} \cdot \mathrm{sinc}(t/T_\mathrm{signal})$. Since $s(t)$ corresponds to the positive Fourier branch, it has to be compared to the two-sided shot noise PSD $q i_\mathrm{LO}$. We therefore substitute $\sigma^2_\mathrm{WN} =q i_\mathrm{LO} \cdot 2B$, and $s(t)$ as defined above into Eq.~\ref{eq:var_tof}. This yields
\begin{align}
    \sigma^2_\mathrm{ToF}(T_0)=\frac{3}{\pi^2}\frac{q T_\mathrm{signal}}{i_\mathrm{signal}}.
\end{align}
This value corresponds to ``oscilloscope time''. To get the corresponding value for optical delay, the variance has to be multiplied by the scale factor $(\Delta f_{\mathrm{rep}}/f_{\mathrm{rep}})^2$. By additionally noting that the measurement time is $T_\mathrm{meas}=1/\Delta f_{\mathrm{rep}}$ and that the full optical bandwidth is given by $\Delta \nu_\mathrm{opt} = 1/T_\mathrm{signal} \cdot f_{\mathrm{rep}}/\Delta f_{\mathrm{rep}}$, we find
\begin{align}
    \sigma^2_\mathrm{ToF, opt}(\tau_0) =& \frac{3}{\pi^2}\frac{1}{f_{\mathrm{rep}} \Delta \nu_\mathrm{opt}} \frac{q}{i_\mathrm{signal} T_\mathrm{meas}}
    \nonumber \\
    =&
    \frac{3}{\pi^2}\frac{1}{f_{\mathrm{rep}} \Delta \nu_\mathrm{opt} N_\mathrm{photon}}
\end{align}
where an effective number of detected signal photons $N_\mathrm{photon}=i_\mathrm{signal} T_\mathrm{meas}/q$ has been introduced. The prefactor $3/\pi^2$ depends on the pulse shape, so for the inequality in Eq.~\ref{eq:CRLB} of the main text we have set it to $O(1)$ to make the result more general.
\bibliography{main}

\end{document}